# On the Relocation Behaviour of Ride-sourcing Drivers

**Peyman Ashkrof, Corresponding author**
*Department of Transport and Planning, Delft University of Technology, Delft, Netherlands*
*Email: P.ashkrof@tudelft.nl*

**Gonçalo Homem de Almeida Correia**
*Department of Transport and Planning, Delft University of Technology, Delft, Netherlands*

**Oded Cats**
*Department of Transport and Planning, Delft University of Technology, Delft, Netherlands*

**Bart van Arem**
*Department of Transport and Planning, Delft University of Technology, Delft, Netherlands*

## Abstract

Ride-sourcing drivers as individual service suppliers can freely adopt their own relocation strategies including waiting, cruising freely, or following the platform recommendations. These decisions substantially impact the balance between supply and demand, and consequently affect system performance. We conducted a stated choice experiment to study the searching behaviour of ride-sourcing drivers and examine novel policies. A unique dataset of 576 ride-sourcing drivers working in the US was collected and a choice modelling approach was used to estimate the effects of multiple existing and hypothetical attributes. The results suggest that ride-sourcing drivers' relocation strategies considerably vary between different groups of drivers. Surge pricing significantly stimulates drivers to head towards the designated areas. However, the distance between driver's location and surge or high-demand areas demotivates them to follow the platform repositioning recommendations. We discuss the implications of our findings for various platform policies on real-time information sharing and platform repositioning guidance.

# 1. Introduction

Ride-sourcing companies - also known as Transport Network Companies (TNCs) – such as Uber and Lyft have been receiving a positive reception from the general public given their growing market share, especially among urban travellers (Conway *et al.* 2018), and have gained more than one-third of the international taxi market (Bryan and Gans 2019). Ride-sourcing is a digital two-sided platform that matches ride requests submitted by riders via a mobile app with available drivers who supply a door-to-door transport service. In this setting, drivers are not only chauffeurs but also private fleet providers. Therefore, ride-sourcing drivers can make various choices at the strategic, tactical, and operational levels. At the operational level, drivers can independently decide on whether to wait around the drop-off location of the last rider, drive to the areas recommended by the platform, or cruise freely with the aim of finding a ride request. This freedom has fundamental implications for the system performance in general and the balance between supply and demand in particular. For instance, the unavailability of drivers in a certain region can increase the rider's waiting time and decrease the match rate, and consequently the system reliability. Furthermore, the so-called idle cruising - referring to moving while no passenger is in the car - can contribute to traffic congestion caused by ride-sourcing operations (Tirachini 2020, Tengilimoglu and Wadud 2021).

Ride-sourcing platforms are interested in steering individual suppliers so as to keep the balance between supply and demand. This is a complex task due to the unpredictable nature of the dynamic demand and the heterogeneity among service suppliers. Platforms adopt various dispatching algorithms, initiatives, and pricing strategies to efficiently reposition empty vehicles and possibly reduce the fleet size and total vehicle mileage. Using taxi trip data in New York, Vazifeh et al. (2018) propose a near-optimal repositioning framework that can decrease the fleet size by 30%. The mainstream of the literature is focused on the optimal algorithms for empty vehicle routing and repositioning to minimize the number of rebalancing vehicles (Zhang and Pavone 2016, Braverman *et al.* 2019) and fleet size (Wen *et al.* 2018, Iglesias *et al.* 2019, Narayan *et al.* 2021), or maximize the profit of the platform and drivers (Godfrey and Powell 2002, Gao *et al.* 2018). Another research direction is concerned with optimal surge pricing as a financial relocation incentive and its implications (Lu *et al.* 2018, Chen *et al.* 2020, Besbes *et al.* 2021). Despite all the proposed approaches and the applied strategies in the real world, there still exist serious challenges and counterproductive results such as a high number of idle vehicles, increasing empty mileage, and traffic congestion (Tirachini 2020, Tengilimoglu and Wadud 2021). Most of the studies assume that the drivers are fully compliant with the repositioning algorithms and policies of a centralized platform and ignore the behavioural aspects of individual drivers. While drivers' strikes worldwide and related court cases demonstrate a widespread dissatisfaction of drivers with the system operations that causes distrust. Such a distrust leads to drivers' dismissal of the platform suggestions and therefore influences the system efficiency and particularly idle repositioning (Özer *et al.* 2018). This calls for gaining a better understanding of drivers' behaviour and their response to various policies and strategies.

There is a growing body of literature aiming to explore the behaviour of ride-sourcing drivers in various aspects (Fielbaum and Tirachini 2020, Xu *et al.* 2020, Zuniga-Garcia *et al.* 2020, Ashkrof *et al.* 2021, He 2021). Ashkrof et al. (2020) carried out a qualitative analysis of system operations from the drivers' perspective and proposed a framework that maps the relationship between the tactical and operational decisions of drivers. They concluded that even though all drivers attempt to maximize their income, their approach differs considerably depending on the platform strategies, drivers' and riders' characteristics, as well as exogenous factors. Analysing 9000 ride-sourcing trips in Beijing, Leng et al. (2016) found out that the idle time of drivers is reduced when a set of financial incentives are offered by the platform. Zuniga-Garcia et al. (2020) demonstrated that the current relocation and pricing algorithms do not sufficiently take drivers' decisions into account. Using trajectory information of the DiDi drivers in China, Xu et al. (2020) reported clear customer search behavioural differences at various time of the day, especially between full-time and part-time drivers. Publicly available ride-sourcing data does not contain, however, information on drivers' positions when travelling without a passenger on-board and therefore cannot fully reveal drivers' repositioning behaviour and preferences. A tailored experiment is therefore needed to investigate the relocation decisions and preferences of drivers under various circumstances.

To the best of our knowledge, this is the first study that is specifically designed to empirically investigate drivers' relocation strategies and their reaction to the platform repositioning guidance. Furthermore, we also study drivers' responses to potential alternative policies and related information provisioned. To this end, a unique dataset of 576 ride-sourcing drivers working in the US is collected using an original carefully designed stated preference survey, and then a choice modelling approach is applied to analyse the data. The findings offer deep insights for platform providers, algorithm developers, policymakers, and other researchers in this field to facilitate the improvement of supply-side operations and planning. The next sections describe the survey design, data collection process, modelling, results, discussion, and conclusions.

## 2. Survey design

Ride-sourcing drivers switch between three repositioning states during their work shift: wait/cruise to find a ride request, drive to pick up an assigned rider, and transport a rider to his/her destination. The first state is primarily dependent on the choices of the individual ride-sourcing driver while the others are mainly directed by the platform. These three states are highly interconnected; therefore, they can influence each other. To illustrate, successful matching, which is the main objective of ride-sourcing systems, is dependent on the availability of idle drivers in proximity to the clients which can be affected by their earlier decisions. Idle ride-sourcing drivers who intend to continue their shift and search for a new ride request have several relocation choices: (i) waiting in a place near the drop-off location of the last fulfilled trip; (ii) following the platform repositioning recommendation (e.g., driving to a surge area or a high-demand area), and; (iii) cruising to move away from the drop-off location neighbourhood based on the driver's experience, preferences, and intuition. Given the inherent difference between surge area, where surge pricing occurs due to a local high imbalance between supply and demand, and high-demand area - locations where the demand is expected to be high while the trip fare remains at the normal rate - driving to surge areas and driving to high-demand areas are considered in the following to constitute two distinctive options.

In this study, we consider the choice situation occurring when the driver has recently completed a ride and is searching for a new passenger while both surge and high-demand areas are available. Therefore, four relocation alternatives are defined:
- Staying as much as possible close to the current location (standstill or driving around)

- Driving to a surge area (shown by a coloured area ranging from light orange to dark red in the app)
- Driving to a high-demand area (marked by a blue flashlight icon in the app)
- Cruising freely into a different area based on the driver's experience, preferences, or intuition

We hypothesize this choice to be dependent on various factors including the spatial-temporal status of drivers, information display settings, driver's working pattern, and their socioeconomics characteristics. To investigate the relocation strategies of ride-sourcing drivers and the explanatory factors, a Stated Choice (SC) experiment is designed. Respondents (ride-sourcing drivers) are asked to choose whether to stay around their current location, follow the surge area, drive to the high-demand area, or cruise freely. The choice is first made based upon a set of existing attributes that drivers currently experience with existing ride-sourcing systems. Subsequently, some currently unavailable information and incentives are added to investigate their potential implications in the relocation choice. Figure 1 illustrates the experiment set-up employed in this study.

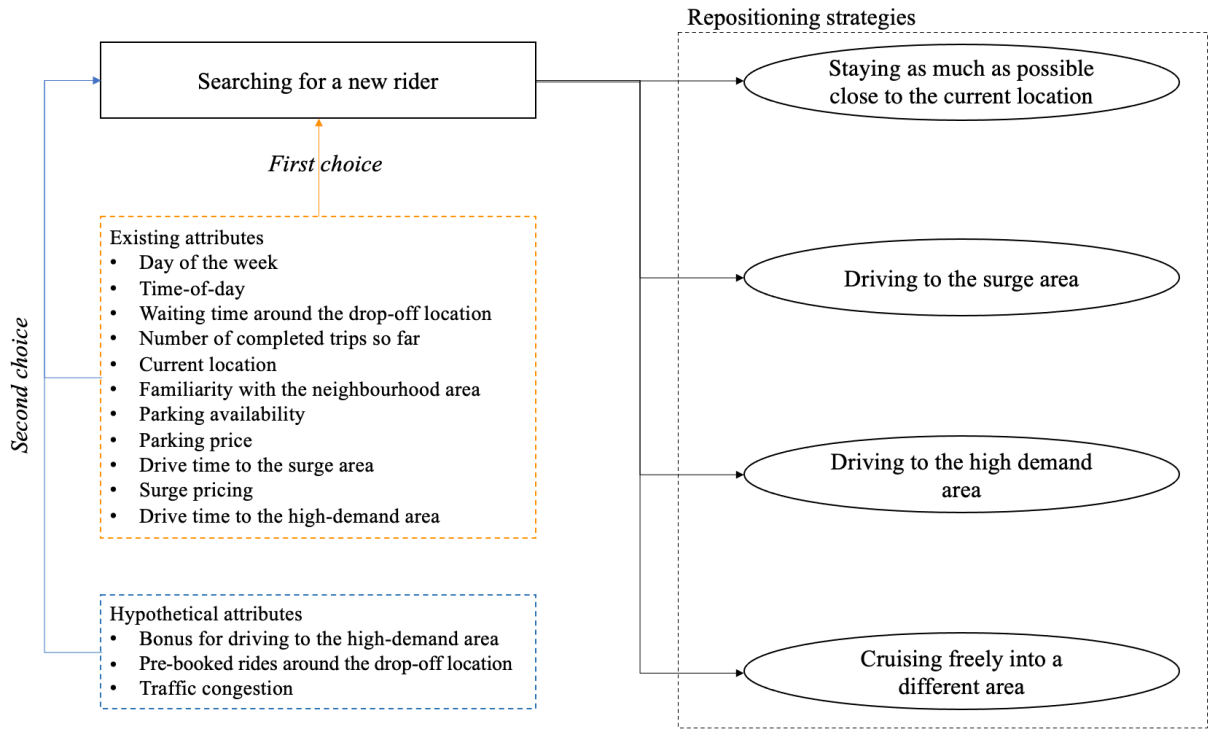

*Fig.1 The stated choice experiment set-up*

All the existing and hypothetical attributes and their respective levels are identified based on the current system operations, driver-side app display, existing literature, interview with drivers (Ashkrof *et al.* 2020), and posts made by drivers on drivers' online forums. Table 1 provides more details about the attributes as well as their respective levels and labels.

*Table 1: Attributes, attribute levels, and labels*

| | Attributes | Definition | Attribute levels/labels |
|---|---|---|---|
| **Existing attributes** | Day of the week | The most common working day | Revealed by the respondent |
| | Time of day | The time that the decision on repositioning is made | Pivoted around the working shift reported by the respondent |
| | Waiting time around the drop-off location [min] | The duration of standing still or driving around the last drop-off spot | 5, 15, 25 |
| | Number of completed trips so far | Number of fulfilled trips since the beginning of the shift | 2, 6, 10, 14 |
| | Current location | The type of operating area | City centre, Suburb |
| | Familiarity with the neighbourhood area | Whether the driver is familiar with the drop-off point area | Familiar, Unfamiliar |
| | Parking availability | Whether a parking spot is available in the vicinity | Available, Unavailable |
| | Parking price [$] | The parking fee in case there is an available parking space | 0, 2, 4 |
| | Surge pricing [$] | A bonus that is offered when the demand is notably higher than the supply | 1, 2, 3 |
| | Drive time to the surge area [min] | Travel time between driver's location and the surge area | 5, 10, 15, 20 |
| | Drive time to the high-demand area [min] | Travel time between driver's location and the high-demand area | 5, 10, 15, 20 |
| **Hypothetical attributes (not currently used by the existing platforms)** | Bonus for driving to the high-demand area [$] | A guaranteed bonus for repositioning to the high-demand area | 1, 2, 3 |
| | Pre-booked rides around the drop-off location [min] | A guaranteed ride if the driver is staying around for the indicated duration at the last drop-off location | 5, 10, 15, 20 |
| | Traffic Congestion | The level of congestion around the drop-off location | Highly congested, Free-flow |

Day of the Week and Time of Day are pivoted around the driver's working pattern. At the beginning of the survey, drivers are requested to state their working days and hours. This information is dynamically used in the survey to create an individual-specific experiment and ensure that drivers can relate to the study context. The Day of the Week is obtained from the respective question and is directly imported to the choice set, while the segmentation technique is applied to determine the levels of Time of Day.

Using this pivot design approach, a library of designs is constructed and respondents are assigned to one of which based on the designated reference point(s). To this end, Time of Day is divided into ten segments based on the driver's shift starting time which can be one of the five time periods (i.e., morning, midday, afternoon, evening, and night) and working duration that can be either a full shift (8 hours) or a half shift (4 hours). Table 2 shows the segmented designs for Time of Day. To illustrate, if a driver starts his/her shift at 10:00 and works for approximately 8 hours, the displayed levels of Time of Day will be 8:00, 12:00, 16:00 for this driver.

*Table 2: The levels of Time of Day pivoted around the driver's working shift*

| Shift starting time | Morning (5:00-11:00) | | Midday (11:00-15:00) | | Afternoon (15:00-19:00) | | Evening (19:00-23:00) | | Night (23:00-5:00) | |
|---|---|---|---|---|---|---|---|---|---|---|
| Working Duration | 8h | 4h | 8h | 4h | 8h | 4h | 8h | 4h | 8h | 4h |
| Time of Day | 8:00 | 8:00 | 13:00 | 13:00 | 17:00 | 17:00 | 21:00 | 21:00 | 2:00 | 2:00 |
| | 12:00 | 10:00 | 17:00 | 15:00 | 21:00 | 19:00 | 1:00 | 23:00 | 6:00 | 4:00 |
| | 16:00 | 12:00 | 21:00 | 17:00 | 1:00 | 21:00 | 5:00 | 1:00 | 10:00 | 6:00 |

To design the SC experiment with a statistically efficient combination of the attribute levels, a Bayesian efficient design is applied. First, the asymptotic variance-covariance (AVC) matrix is estimated by calculating the negative inverse of the expected second derivative of the loglikelihood function of the choice model. Subsequentially, the standard error of the parameter estimates is obtained from the roots of the diagonal of the AVC matrix and then is minimized to find an efficient design measured by an efficiency measure. The most widely used efficiency measure is the so-called D-error which is the determinant of the AVC matrix (Bliemer and Rose 2010). Given that no prior knowledge about the parameter estimates is available, the design was initially constructed using $D_z - error$ assuming the priors equal to zero (orthogonal):

$$D_z - error = \det(\Omega\ (X, 0))^{1/K} \qquad \text{Eq. (1)}$$

Where $\Omega$ denotes the AVC matrix, $X$ is the choice set design, and $K$ refers to the number of parameters. Then, a pilot of 50 responses was conducted to estimate the priors and construct the AVC matrix. To achieve a more reliable design that is less dependent on the exact priors, the Bayesian design is used. In this method, the priors are assumed to be random variables expressing the uncertainty about the parameter value. To this end, the so-called $D_z - error$ expressed in Eq. (2) is used:

$$D_z - error = \int \det(\Omega_1\ (X, \tilde{\beta}))^{\frac{1}{K}}\ \emptyset(\tilde{\beta}|\theta)d\tilde{\beta} \qquad \text{Eq. (2)}$$

Where $\tilde{\beta}$ is a random variable with a joint probability distribution function $\emptyset$ given parameter $\theta$. In this study, $\tilde{\beta}$ is assumed to be uniformly distributed: $\tilde{\beta}(u, v)$ where $u$ and $v$ are the mean and standard deviation, respectively, obtained from the pilot phase. The software package NGENE (ChoiceMetrics 2018) was used to construct 24 choice sets in 6 blocks that were randomly distributed between respondents.

A survey software platform is used to program an online questionnaire that enables the data collection process. To make sure that respondents comply with the survey requirements (i.e., being an active ride-sourcing driver working at least once a week), a series of screening questions is deployed at the beginning of the questionnaire. Eligible drivers are asked to provide details of their working pattern which then, as explained above, feed the segmented design. Next, the introduction to the choice experiment coupled with an example is shown and then respondents are requested to indicate their relocation choices based on the information provided. Figure 2 provides an illustration of the choice set displayed in each scenario. The last section in the survey collects respondent-specific information such as the driver's working as well as socio-demographic characteristics.

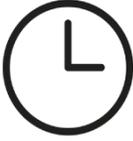

**Fig.2** *Choice set interface with the existing (left) and hypothetical (right) attributes*

## 3. Survey data

In this study, Uber and Lyft drivers working in the United States were selected to be part of the survey sample. A panel company was hired to recruit prospective respondents for this hard-to-reach target group. In total, 752 complete responses were collected between November 2020 and February 2021. A comprehensive data quality analysis was performed to filter out low-quality responses caused by short response time and the lack of sufficient attention. As a result, 576 responses were retained for the analysis.

The descriptive statistics of the data show that more than 85% of the drivers drive for Uber while only about 15% of them work solely with Lyft. Around 40% of the drivers are fully financially reliant on the ride-sourcing job, labelled as full-time drivers. These drivers also work on average more hours per week than part-time drivers - drivers who have other employment-related income. Regarding work experience, most of the drivers have been working as ride-sourcing drivers for the last 13-36 months. The most common workday is Monday. Furthermore, more than 70% of the drivers work in the morning shift for either 4 or 8 hours. About 70% of the sample consists of male drivers and more than 80% of the drivers are younger than 40 years old.

## 4. Discrete Choice Modelling

A discrete choice modelling approach is applied to unravel the relocation strategies of drivers and identify the influential existing and potential factors. Assuming that both surge and high-demand areas are available, we define four choice alternatives: waiting around, driving to a surge area, heading to a high-demand area, and cruising freely based on their experience and intuition. Then, the identified attributes are used to formulate the utility function of alternative $j$ as follows:

$$U_j = \sum_{k=1}^{K} \beta_{jk} \cdot x_{jk} + \sum_{m=1}^{M} \beta_m \cdot x_m \qquad \text{Eq. (3)}$$

Where the first term refers to the alternative-specific attributes ($x_{jk}$) presented in the choice experiment, the second component includes the individual-specific factors such as driver's socio-economic characteristics ($x_m$), and the last component is the error term ($\varepsilon_j$) that captures the unexplained variation under the assumption of being independently and identically distributed. $\beta_{jk}$ and $\beta_m$ are the coefficients vectors representing the marginal effects of the exploratory attributes and individual-specific factors respectively. The Random Utility Maximation (RUM) approach is used to estimate the choice models by the software package PandasBiogeme (Bierlaire 2020).

## 5. Model estimation results

A bottom-up modelling approach is applied to estimate various models with possible meaningful inclusion of the attributes into the utility functions. Based on the incorporated variables, four models divided into two groups are reported. At the upper level, two scenarios are defined based on the information shown to drivers (existing and hypothetical). For each scenario, two models are estimated distinguished by the variables incorporated into the DCMs:

- Primary: This model contains solely the variables displayed in the choice experiments.
- Full: The working and socio-demographic characteristics of the drivers are added to this model.

This incremental inclusion of categories of variables enables the understanding of the impacts of different types of attributes on the repositioning decision of ride-sourcing drivers depending on the application of interest and the available information. For example, in a future application of the choice model in case that no information about the characteristics of individual drivers is available, the primary model can still be used.

Table 3 shows the results of the models built upon the existing and hypothetical attributes. ASC represents the alternative specific constant, the suffixes W (Waiting/staying around), S (driving to the Surge area), H (driving to the High-demand area), and C (Cruising freely) indicate the utility function for which the attribute is relevant.

*Table 3: The results of the choice models built upon the existing and the hypothetical attributes*

| Parameters | Scenario 1 (only existing attributes) | | Scenario 2 (with hypothetical attributes) | |
|---|---|---|---|---|
| | **Primary** | **Full** | **Primary** | **Full** |
| ASC_Waiting | 0.207 | -0.283 | 1.12*** | 0.988*** |
| Waiting Time_W [min] | -0.022*** | -0.019** | -0.019** | -0.017** |
| Number of Trips_S&H | 0.080*** | 0.060*** | 0.075*** | 0.048*** |
| Driver's Location_W [1=City center] | 0.322** | 0.315** | -0.009 | -0.022 |
| Familiarity with Neighborhood_C [1=Familiar] | -0.312** | -0.201 | -0.443*** | -0.334* |
| Parking Availability_W [1=Available] | 0.286** | 0.277** | 0.363*** | 0.325** |
| Surge Pricing_S [$] | 0.177*** | 0.190*** | 0.165*** | 0.166*** |
| Drive Time to Surge Area_S [min] | -0.020*** | -0.020*** | -0.016* | -0.017* |
| Drive Time to High-Demand Area_H [min] | -0.025*** | -0.037*** | -0.035*** | -0.042*** |
| Working on Weekend/Friday_W | 0.350*** | 0.427*** | 0.183* | 0.236* |
| Working Shift_C [1=Beginning of the shift] | -0.764*** | -0.583*** | - | - |
| Beginners_W&C [1=Beginners] | - | -0.322** | -. | -0.018 |
| Part-time Drivers_W [1=Part-time] | - | 0.393*** | - | - |
| High Acceptance Rate_W [1=Acceptance rate>70%] | - | -0.407*** | - | -0.369*** |
| Fully Satisfied Drivers_H [1=Fully satisfied] | - | 0.371*** | - | 0.524*** |
| Taxi Driving Experience_C [1=Taxi driver] | - | -0.478*** | - | -0.370** |
| Educated Driver_W [1=Educated] | - | 0.406*** | - | 0.003 |
| Working Shift_W [1=Beginning of the shift] | - | - | -0.476*** | -0.344** |
| Part-time Drivers_C [1=Part-time] | - | - | - | -0.326** |
| Pre-Booked Rides_W [min] | - | - | -0.021* | -0.020* |
| Bonus to Drive to High-Demand Area_H [$] | - | - | 0.264*** | 0.177*** |
| Traffic Congestion_C | - | - | -0.407** | -0.283* |
| **Initial Log-Likelihood** | -3194.022 | -3194.022 | -3194.022 | -3194.022 |
| **Final Log-Likelihood** | -2991.151 | -2938.735 | -2986.493 | -2949.844 |
| **Rho-square** | 0.064 | 0.080 | 0.065 | 0.076 |

*Significance code: \*p-value<0.05, \*\*p-value<0.01, \*\*\*p-value<0.001*

We first review the results of the models estimated for the current information display setting and then proceed with reporting the results of the hypothetical scenario. The negative value of Waiting Time_W suggests that drivers tend to move to a different area in case the waiting time around the drop-off location increases. On the other hand, drivers working on weekends as well as Fridays are inclined to wait around their location. This might stem from the relatively higher demand on these days of the week (Rangel *et al.* 2021). Therefore, drivers can receive more requests with less driving effort (operational costs). Based on the current system setting, at the beginning of the shift, there is a strong aversion to cruise freely. This might be because the risks of self-determining movements are typically higher, therefore, drivers are willing to first try out waiting or following the platform's suggestions. Interestingly, drivers who have had the experience of being conventional taxi drivers prior to joining the platform dislike cruising on their own and have a tendency to chase the platform repositioning recommendations (i.e., high-demand/surge area) or stay at a particular location to receive a ride request. This could be attributed to their past experience in cruising as taxi drivers, leading them to opt for a system that offers more guidance.

The number of completed trips since the beginning of the shift has a positive effect on driving to the surge and high-demand areas. A satisfactory working experience can develop trust between drivers and the platform which leads to a higher willingness to follow the app recommendation. This is in line with the positive significant value of Fully Satisfied Drivers_H that suggests that highly satisfied drivers (i.e., the drivers who gave 4.5/5 out of 5 stars to the system performance) are more likely to drive to a high-demand area indicated by the platform. Moreover, the results of the first scenario suggest that beginning drivers with a working experience of less than one year (most of whom have high trust in system operations) prefer not to wait or cruise freely but drive to the surge and high-demand areas.

The chance of staying close to the current location is higher in the city centre where the probability of receiving a ride while standing still or driving around is higher compared to a suburban area. Parking availability is also a crucial factor that motivates drivers to wait at a particular location to receive a new ride request. Another influential determinant is the employment status of drivers. Part-time drivers tend to stay around. They need to minimize their operational costs during their working time which is limited by other working activities. That is why they might be more reluctant to move into new areas. Drivers who have a college degree or higher are also more inclined to wait, everything else being the same. We also examine the relation between ride acceptance behaviour and repositioning strategy. We find that drivers with an acceptance rate of more than 70% tend to move as opposed to waiting. These drivers are less selective in assessing ride requests and their intention is to find a ride as quickly as possible, paying less attention to its attractiveness.

As expected, surge pricing stimulates drivers to head to the surge area as they can expect to earn more money in the case of reaching the designated area, receiving and accepting a ride request within the surge pricing period. On the other hand, a higher distance to a surge or a high-demand area discourages drivers to follow the platform repositioning suggestions. This is because the demand-supply intensity dynamically changes and the risk of missing the opportunity is higher when the distance increases. The value of drive to the surge area which is the amount of surge pricing for every minute added to the travel time to the surge area is estimated to be roughly 0.11 $/min based on the results of the Primary model.

When drivers are provided with more information and incentives, some new alternative-specific factors start playing an essential role while the impact of some existing variables change.

Moreover, even several attributes such as the driver's location, experience level, and education are no longer statistically significant at the 95% level. A strong unobserved preference for staying around is identified in the second scenario. Moreover, being familiar with the neighbourhood area increases the probability of waiting or driving to the surge or high-demand area. Presumably, this familiarity helps drivers to find suitable spots to wait or choose the best route to promptly reach the surge or high-demand area.

The existence of pre-booked rides around the drop-off location can influence the choice of drivers to stay around. This hypothetical attribute gives drivers information about the next potential client who can be picked up within their current zone. If drivers declare their interest in waiting for the incoming request, the ride will be secured for them. Nevertheless, drivers may prefer not to stay if the waiting time is relatively high. Moreover, drivers are more likely not to wait at the beginning of the shift arguably because alternative promotions including surge pricing and high-demand bonus can be expected.

Another variable included in the second scenario is the bonus for driving to a high-demand area. The positive significant value of the estimated parameter suggests that drivers are highly inclined to reach the high-demand area if a promotion is offered. Drivers are about 60% more sensitive to the high-demand bonus than towards surge pricing. This is because unlike surge pricing which is paid only if a rider is picked up, this bonus is guaranteed if the driver is driving towards the high-demand area. This has a potential implication when the platform intends to redistribute the available fleet, especially when drivers do not deliberately follow the surge area.

Traffic congestion around the current location turns out to be a significant determinant. A highly congested area discourages drivers to cruise freely given that they probably get stuck in the traffic congestion without picking up passengers – increasing the operational costs. Due to the more restricted time, part-time drivers are less inclined to cruise freely and are more responsive to financial promotions and extra information offered by the platform than full-time drivers, everything else being equal.

## 6. Discussion and conclusions

We empirically study the relocation behaviour of ride-sourcing drivers. To this end, we designed a stated choice experiment to allow investigating the behaviour of drivers under the existing system settings as well as under a hypothetical scenario exploring their potential responses in the event of new circumstances. In total, 576 qualified responses from Uber and Lyft drivers working in the United States were collected, and a series of discrete choice models were estimated. Four choice alternatives were considered: staying around the drop-off location, driving to a surge area, driving to a high-demand area, and cruising freely. Indicating surge and high-demand areas are the most well-known examples of platforms' repositioning guidance. Moreover, various existing and hypothetical incentives and information about driving conditions and demand characteristics were shared with drivers to identify the influential determinants and their potential effects. We also investigated the impacts of other aspects of driver's behaviour at the tactical level (working shift) and the operational level (ride acceptance behaviour) as well as other individual attributes.

Surge pricing - also known as dynamic pricing - is an incentive offered by platforms to redistribute the available fleet and address local imbalances in supply-demand ratios. Platforms also indicate high-demand areas where demand is relatively high but without changes to the normal rate (for both riders and drivers). In general, platform repositioning guidance is a controversial policy that provokes serious disputes. On one hand, Jing et al. (2018) and Jiao (2018) argue that the unpredictability and ambiguity of surge pricing harbour serious doubts among drivers. On the other hand, surge pricing is considered to be a near-optimal solution that can increase the match rate as well as drivers' income (Cachon *et al.* 2017, Lu *et al.* 2018, Nourinejad and Ramezani 2019, Ashkrof *et al.* 2021). Conducting a focus group study with Uber drivers, Ashkrof et al. (2020) reported that some drivers, in particular experienced ones, distrust surge pricing as well as high-demand areas and do not follow them. Those drivers believed that the platform misleads them by not reporting surge and high-demand areas in real-time in order to relocate them to a particular location. These are in line with our findings that suggest that following the surge and high-demand area appears to be more attractive for some groups of drivers depending on their working experience, operational performances, and satisfaction level. Namely, relatively inexperienced drivers, as well as highly satisfied drivers, and drivers with a higher number of completed trips since the beginning of their shift are more likely to follow the recommended areas. The level of surge pricing and the expected travel time between the driver's location and the surge/high-demand area are recognized as the other significant determinants.

Additional repositioning guidance options which are not yet available were studied in the hypothetical scenario. Drivers were given some additional information including the existence of any pre-booked rides in the waiting area (associated with the waiting alternative), bonus for driving to the high-demand area, and the level of congestion around their location (which may impact propensity for cruising freely). We found all these variables can play a role in the relocation choice of drivers. Pre-booked rides can be shown to drivers in advance to enable them to assess whether to stay or not depending on the expected waiting time. In order to motivate drivers to relocate to a particular area such as a high-demand area, a guaranteed bonus may be offered. This guaranteed bonus is valued 60% more highly than surge pricing which is not necessarily secured. Obviously, the platform will need to determine how to set such a bonus in a way that is beneficial for its operations. These results suggest that platform guidance policy may be extended to assist drivers in making more informed decisions and thus possibly improve the level of service, reduce deadhead movements, and improve the wider acceptability of ride-sourcing services.

Our findings can be used to consider the underlying determinants of drivers' behaviour in predicting their relocation choices and designing tailored drivers' incentives. For instance, educated part-time drivers with low acceptance rate who are more likely to stay around can be provided with more information about available parking spots and pre-booked rides in the vicinity, especially when working in the city centre on weekends and Fridays. In contrast, beginning drivers are more willing to respond to detailed information about surge and high-demand areas. This is in line with the findings of Tengilimoglu and Wadud (2021) that acknowledge the behavioural heterogeneity among drivers and conclude that a more effective management is needed to reduce the empty mileage of ride-sourcing services. Given that trust between individual suppliers and the platform is key in the success of such an interactive business model (Özer *et al.* 2018), the information shared by the platform needs to be accurate and unbiased and communicated in real-time to build the basic trust and develop it over time.

The results of this study can also be used as input to ride-sourcing simulation models to include the relocation behaviour of drivers, explore various policy designs, and investigate their impacts on system operations. Future research may validate the results of this study using revealed preference data.

## Acknowledgement

This research was supported by the CriticalMaaS project (grant number 804469), which is financed by the European Research Council and the Amsterdam Institute for Advanced Metropolitan Solutions.